\documentclass{article}

\usepackage{PRIMEarxiv}

\usepackage{multirow}

\usepackage[utf8]{inputenc} 
\usepackage[T1]{fontenc}    
\usepackage{hyperref}       
\usepackage{url}            
\usepackage{booktabs}       
\usepackage{amsfonts}       
\usepackage{nicefrac}       

\usepackage{microtype}      
\usepackage{lipsum}
\usepackage{fancyhdr}       
\usepackage{graphicx}       

\graphicspath{{media/}}     

\title{DERM12345: A Large, Multisource Dermatoscopic Skin Lesion Dataset with 38 Subclasses
}

\author{
  Abdurrahim Yilmaz$^1$ \\
   \And
  Sirin Pekcan Yasar$^2$ \\
   \And
   Gulsum Gencoglan$^{3,\dag}$ \\
  \And
   Burak Temelkuran$^{1,\dag}$ \\
}

\begin{document}
\maketitle

\centerline{\begin{minipage}{13cm}
{1. Department of Metabolism, Digestion, and Reproduction, Imperial College London,  London, SW7 2AZ, United Kingdom\\}
{ 2. Department of Dermatology, The University of Health Sciences, Haydarpasa Numune Research and Training Hospital,  Istanbul, 34668, Turkiye\\}{ 3. Department of Dermatology, Istinye University, Liv Hospital Vadistanbul,  Istanbul, 34010, Turkiye\\}{ Corresponding authors: Gulsum Gencoglan (ggencoglan@gmail.com), Burak Temelkuran (b.temelkuran@imperial.ac.uk)\\}
\end{minipage}}

\begin{abstract}
Skin lesion datasets provide essential information for understanding various skin conditions and developing effective diagnostic tools. They aid the artificial intelligence-based early detection of skin cancer, facilitate treatment planning, and contribute to medical education and research. Published large datasets have partially coverage the subclassifications of the skin lesions. This limitation highlights the need for more expansive and varied datasets to reduce false predictions and help improve the failure analysis for skin lesions. This study presents a diverse dataset comprising 12,345 dermatoscopic images with 38 subclasses of skin lesions collected in Türkiye which comprises different skin types in the transition zone between Europe and Asia. Each subgroup contains high-resolution photos and expert annotations, providing a strong and reliable basis for future research. The detailed analysis of each subgroup provided in this study facilitates targeted research endeavors and enhances the depth of understanding regarding the skin lesions. This dataset distinguishes itself through a diverse structure with 5 super classes, 15 main classes, 38 subclasses and its 12,345 high-resolution dermatoscopic images.
\end{abstract}


\section*{Background \& Summary}
Dermatoscopy, also known as epiluminescence microscopy, is the inspection of the skin using magnifying lenses and polarized or non-polarized filtered illumination, allowing examination of lesions invisible to the naked eye. It is the most extensively used and standardized diagnostic procedure in clinics to diagnose skin lesions. Various types of dermatoscopes are actively used in daily clinics. Hand-held dermatoscopes, for example, have become widely available, affordable, and widely employed, especially in developing countries. However, dermatologists' diagnostic performance in dermatoscopic skin lesion identification is closely linked to their training and prior clinical experience, highlighting a variability inherent in human decisions. The prevalence and structure of malignant and benign lesions that vary depending on race, geographical factors and skin type also affects dermatologists’ performance. Addressing this gap, artificial intelligence (AI) trained on specialized dataset can reduce the experience gap that results from the human factor during the clinical examination of these lesions \cite{tschandl2020human}.

One of the initial studies was published on the classification of only three classes: melanoma, nevus, and seborrheic keratosis \cite{gutman2016skin, codella2018skin}. The skin lesion dataset HAM10000, which had seven classes and common lesions, was published in 2018 \cite{tschandl2018ham10000}. Eight classes with the squamous cell carcinoma skin lesion which is a common lesion but uncovered in HAM10000 was later presented by the BCN20000 dataset \cite{combalia2019bcn20000}. Among dermatoscopic skin lesion datasets, BCN20000 has the largest number of multi-class images (eight classes), still requires further development and expansion to include subclasses. New datasets are continued to be published in open access repositories such as The International Skin Imaging Collaboration (ISIC), which currently has a total of 76,108 public images (accessed at: 30 May 2024) \cite{ricci2023dataset,marchetti2023prospective}. Publishing new datasets enables the development of more robust and intelligent AI algorithms with different approaches.

Datasets on skin lesions can be listed in four different categories: clinical, pathological slide, and dermatoscopic image datasets, and datasets combining more than one of these modalities. Clinical image datasets such as the Interactive Atlas of Dermatoscopy and the Dermofit Image Library can be accessible for a fee \cite{argenziano2000interactive, dermofit}. The MED-NODE Dataset \cite{giotis2015med}, the Asan and Hallym Dataset \cite{han2018classification}, the SD-198/SD-260 Datasets \cite{sun2016benchmark, yang2018clinical, yang2019self}, PAD-UFES-20 \cite{pacheco2020pad}, SCIN (5.2\% skin lesion) \cite{ward2024crowdsourcing}, Atlas Dermatologica \cite{Atlasdermatologico}, and DermaAmin \cite{dermaamin} are all open-access. Fitzpatrick 17k Dataset is a dataset sourced from Atlas Dermatologica and DermaAmin, with the addition of Fitzpatrick skin type information (28.62\% skin lesion) \cite{groh2021evaluating}. Diverse dermatology images (DDI) dataset also presents skin images with their skin type \cite{ddistanford}. Cancer Genome Atlas is a pathological slide image dataset for skin lesions in the literature \cite{genome}. Clinical, dermatoscopic, and pathological images can all be found on Dermnet NZ for a free (high-resolution images for a fee) \cite{dermnet}. Study results show that the diagnostic accuracy (by clinicians and AI) achievable using dermatoscopic image datasets result in higher success rates when compared to those achievable using clinical image datasets, bringing out the importance of the dermatoscopic image datasets \cite{yap2018multimodal}. Pathological images require tissue excision from the patient. For that reason, pathological image dataset is not as practical as dermatoscopic observation.

PH$^2$ dataset was published for benchmarking dermatoscopic images including with three lesion classes (common nevus, atypical nevus and melanoma (mel)) \cite{mendoncca2013ph}. The Derm7pt dataset with five lesion classes (basal cell carcinoma (bcc), mel, miscellaneous, nevus (nv), and seborrheic keratosis (sk)), 2000 dermatoscopic and clinical images was published along with the 7-point checklist pattern analysis and its automation by using artificial neural networks \cite{kawahara2018seven}. The largest dermatoscopic dataset included in the scientific studies is the International Skin Imaging Collaboration (ISIC) archive with 240,000+ total images (accessed at: 30 May 2024). This archive contains these datasets: ISIC-2016 with 1,279 images and two lesion classes (benign and malignant) \cite{gutman2016skin}, ISIC-2017 with 2,000 images three lesion classes (mel, nv, and sk) \cite{codella2018skin}, ISIC-2018 \cite{codella2019skin} and HAM10000 with 10,015 images and seven lesion classes (akiec, bcc, benign-keratosis like (bkl), dermatofibroma (df), mel, nv, and vascular (vasc)) \cite{tschandl2018ham10000}, BCN20000 with 19,424 images and eight lesion classes (actinic keratosis (ak), bcc, df, mel, nv, squamous cell carcinoma (scc), sk, and vasc) \cite{combalia2019bcn20000} and Patient-Centric dataset with 33,126 images and two lesion classes (benign  and mel) \cite{rotemberg2021patient}. Lastly, the dataset collected in Argentina with 10 subclasses (ak, bcc, df, mel, nv, scc, sk, solar lentigo (sl), lichenoid keratosis (lk), and vasc) was added into ISIC dataset \cite{ricci2023dataset}. Between 2016 and 2020, numerous competitions were held on challenges such as lesion segmentation, feature extraction, and lesion classification, using these ISIC datasets.

These large datasets are crucial for the development of AI based models for skin lesion classification purposes. In addition, the introduction of subclass annotations in skin image datasets has high potential to enhance studies in the AI field focusing on skin disease, for creation of more trustworthy, robust, and intelligent systems. Here, our study presents a taxonomic tree for skin lesion classes and a dataset of 38 skin lesions annotated with subclasses for the first time in the scientific literature. This dataset, which contains a total of 12,345 images, is a comprehensive dermatoscopic image dataset available. Moreover, this is the one of the largest collection to date in terms of the total number of images in multiclass datasets.

\section*{Methods}
\begin{figure*}
\setlength{\fboxsep}{0pt}%
\setlength{\fboxrule}{0pt}%
\centering
\includegraphics[width=\textwidth]{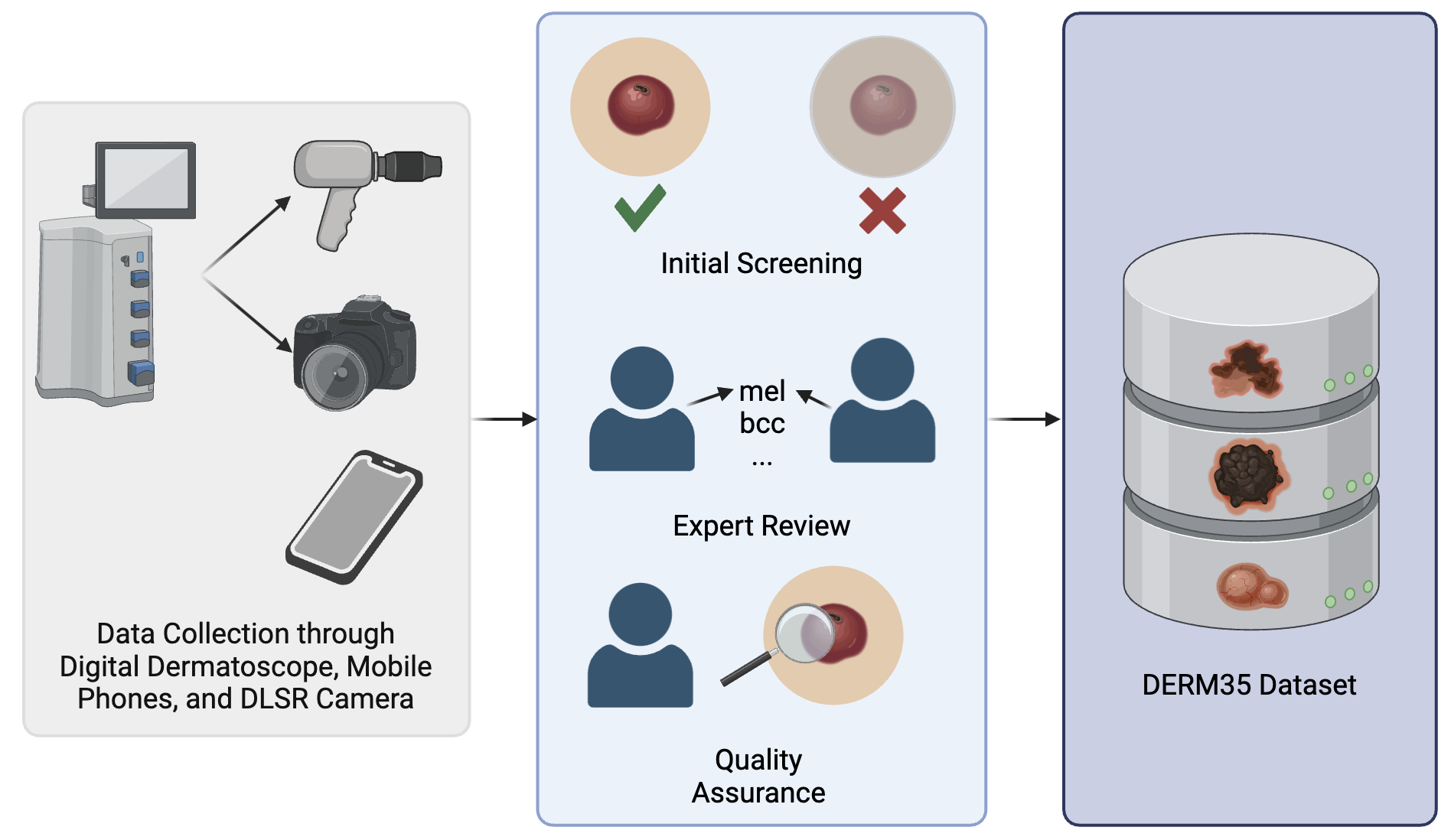}
\caption{Shows an overview of the data collection procedure.}
\label{fig:overview}
\end{figure*}

The DERM38 dataset contains 12,345 high-resolution dermatoscopic images from 1,761 patients, which were collected from 2008 to 2020 in the Department of Dermatology and Venerology at Celal Bayar University (Manisa, Türkiye), Istinye University (Istanbul, Türkiye) and University of Health Sciences Haydarpasa Numune Research and Training Hospital (Istanbul, Türkiye) by using MoleMax 3, MoleMax HD (Derma Medical Systems, Vienna, Austria), FotoFinder® videodermatoscope (FotoFinder Systems, Bad Birnbach, Germany) and 3gen Dermlite DL4 hand-held dermatoscopes (DermLite LLC, California, United States of America) with connection kit for iPhone 5 and 7 (Apple Inc., California, United States of America). Ethical approval of images was based on ethics review board protocols 20.478.486/1023 (Manisa Celal Bayar University, 24/11/2021). The informed consent was waived because of the retrospective nature of the study as the dataset contains anonymized data. The final dataset includes 38 skin lesion classes. The collection process includes capturing a wide variety of skin lesions using both digital dermatoscopy devices and high-resolution digital single-lens reflex (DSLR) cameras with dermatoscopy attachment. The goal is to improve a diverse representation of data in terms of image quality, magnification, and resolution. The overview of data collection is shown in Fig. \ref{fig:overview}.

\subsection*{Data collection from digital dermatoscopy devices}
Digital dermatoscopy devices are computer-attached systems and outfitted with dermatoscopes that utilize polarized light, enabling a magnification such as from 10x to 140x. This feature facilitates the thorough examination of deeper skin structures with a high level of detail. These dermatoscopy devices are linked to a computer system equipped with specialized software for the purpose of capturing and storing images. By using this software, dermatologists can follow up on the patients and their skin lesions with detailed information such as location and lesion classes. Each lesion is captured using a standardized technique and consistent lighting conditions to minimize any potential variability. Dermatoscopy is performed in the contact-polarized mode using an interface medium to avoid reflection caused by excessive scales, with ultrasound gel applied when necessary. The dermatologists ensure the most possible and accurate alignment of the field of view, with the lesion at the center and a small margin of surrounding healthy skin included to provide contextual information. To generate our dataset, we exported the data using their software tools in suitable formats such as HTML files. The raw data was subsequently retrieved from the raw files as lesion images and their metadata information. The cases were selected from this system by two expert dermatologists according to their consensus benign diagnosis, follow-up, and excised lesions with a histopathologic report. 

\subsection*{Data collection from mobile phones and hand-held dermatoscopes}
Mobile phone integrated dermatoscopes are low-cost and accessible devices that can provide state-of-art image quality. In this dataset, Apple iPhone 5 and 7, equipped with a 3gen Dermlite DL 4 hand-held dermatoscope kits, were used to collect dermatoscopic images. The Dermlite DL4 attachment with mobile devices enables the utilization of polarized light dermatoscopy, which offers a maximum magnification of 10x. The use of high-level magnification is essential for thoroughly analyzing the intricate components of the skin. The mobile phone was utilized alongside dermatoscope attachments, facilitating the acquisition of high-resolution (such as at 3840x2160 pixels, 4K resolution) skin lesion photos directly onto the phone. The mobile photos, along with their metadata, including lesion classes, were systematically documented, and matched with their records on digital dermatoscopy devices. A standardized image acquisition process was adhered to keep consistent lighting and framing conditions for all photos. The dermatologists conducted the imaging by positioning the lesions in the center of the field of view as possible while also including a border of healthy skin to offer contextual information. When focusing or centering the mobile phone camera was challenging, the dermatologists took photos in different angles.  This usually caused the edges of the dermatoscopes as a large black region to be visible on these images. These black regions were cropped from the images meticulously to increase standardization of images.

\subsection*{Data selection and quality control}
The raw dataset included not only dermatoscopic images but also non-dermatoscopic images and other images such as device test images. To accelerate the data selection procedure, the dermatoscopic images were extracted by utilizing a script in Python (version: 3.11.5) that selected the images together with their metadata information. The BeautifulSoup library (version: 4.12.2) was utilized for the automated extraction and categorization of photos in HTML format. Data that could not be extracted or exported automatically in HTML format was hand-labeled and classified. The metadata was additionally stored manually in comma-separated values (CSV) files. To remove duplicate images, their file sizes were compared by using a script coded in Python. These scripts found 280 duplicate images. Then, they were manually reviewed. Consequently, all dermatoscopic images were retrieved and their corresponding information was stored in a CSV file. After automated extraction of the images, the careful selection of data was facilitated to maintain the integrity and usefulness of the dataset for both clinical reference and computational purposes. The criteria for selection were determined prior to the initiation of data collection and were consistently followed during the selection procedure.

\textbf{Inclusion Criteria}
\begin{itemize}
\item \textbf{Image Quality:} Images must be clear, in focus, and have sufficient quality. All images were reviewed manually to ensure that key diagnostic features of the skin lesions are understandable. 
\item \textbf{Diagnostic Confirmation:} Only images of lesions with a consensus of clinical diagnosis by two expert dermatologists or a histopathological confirmation (where available) were included.
\end{itemize}

\textbf{Exclusion Criteria}
\begin{itemize}
\item \textbf{Poor Image Quality:} Images that were blurred, under- or over-exposed, or had large artifacts. All images were reviewed manually, and the images that could potentially interfere with diagnosis were excluded.
\item \textbf{Incomplete Data:} Images lacking essential metadata such as the presumed diagnosis were not included in the dataset.
\item \textbf{Ethical Concerns:} Images that contained identifiable patient information or did not have proper consent were excluded to uphold ethical standards.
\item \textbf{Unsuitable:} Images which contain unsuitable skin diseases or artefacts were excluded. Nail lesions, mucosal lesions, cockade nevus, leukoplakia, sarcoidosis psoriasis, verruca, cyst, keloid, sebaceous hyperplasia, wart, genital wart, lipoid proteinoz, subcorneal hemorrhage, pigmented purpuric dermatosis, mucosal melanosis, pygmented bowen, verruca, sarkoidoz psoriasis, and some types of adenocarcinoma such as sebaceous carcinoma and eccrine glands were excluded.
\end{itemize}

\textbf{Selection Process}
The selection process involved multiple stages:
\begin{itemize}
\item \textbf{Initial Screening:} A preliminary review of images was conducted by trained engineers to remove images that obviously did not meet the inclusion and exclusion criteria. If one more than different classes of skin lesion is located in an image, these images are cropped.
\item \textbf{Expert Review:} Two expert dermatologists then reviewed the remaining images to ensure that they met the clinical standards and diagnostic requirements.
\item \textbf{Quality Assurance:} A final review was conducted by a panel comprising two dermatologists (GG and SPY with over 20 years experience on dermatoscopy) to ensure the images were of suitable quality for both clinical reference and computational tasks.
\end{itemize}

\begin{table}
\centering
\caption{The DERM38 dataset contains 12,345 skin lesions divided into 5 super classes, 15 main classes, and 38 subclasses.}
\label{tab:class_results}
\begin{tabular}{l|l|l|l}
Superclass                                & Mainclass                              & Subclass                        & \# of Images  \\ 
\hline
\multirow{19}{*}{Melanocytic Benign}      & \multirow{4}{*}{Banal Compound}        & Acral (an)                          & 22            \\
                                          &                                        & Congenital (con)                     & 423           \\
                                          &                                        & Miescher (mn)                       & 38            \\
                                          &                                        & Nevus (cn)                          & 673           \\ 
\cline{2-4}
                                          & \multirow{2}{*}{Banal Dermal}          & Blue (bn)                            & 156           \\
                                          &                                        & Nevus (dn)                          & 539           \\ 
\cline{2-4}
                                          & \multirow{3}{*}{Banal Junctional}      & Acral (an)                          & 351           \\
                                          &                                        & Congenital (con)                     & 145           \\
                                          &                                        & Nevus (jn)                          & 1419          \\ 
\cline{2-4}
                                          & \multirow{3}{*}{Lentigo}               & Ink Spot Lentigo (isl)               & 6             \\
                                          &                                        & Lentigo Simplex (ls)                & 23            \\
                                          &                                        & Solar Lentigo (sl)                  & 65            \\ 
\cline{2-4}
                                          & \multirow{3}{*}{Dysplastic Compound}   & Acral (an)                          & 8             \\
                                          &                                        & Congenital (con)                     & 30            \\
                                          &                                        & Nevus (cn)                          & 426           \\ 
\cline{2-4}
                                          & \multirow{3}{*}{Dysplastic Junctional} & Acral (an)                          & 212           \\
                                          &                                        & Spitz (sn)                          & 10            \\
                                          &                                        & Nevus (jn)                          & 5465          \\ 
\cline{2-4}
                                          & Dysplastic Recurrent                   & Recurrent (rn)                      & 15             \\ 
\hline
\multirow{5}{*}{Melanocytic Malignant}    & \multirow{5}{*}{Melanoma}              & Acral Nodular (anm)                  & 78            \\
                                          &                                        & Acral Lentiginious (alm)              & 54            \\
                                          &                                        & Lentigo Maligna (lm)                & 85            \\
                                          &                                        & Lentigo Maligna Melanom (lmm)        & 21            \\
                                          &                                        & Melanoma (mel)                       & 164           \\ 
\hline
\multirow{6}{*}{Nonmelanocytic Benign}    & \multirow{2}{*}{Keratinocytic}         & Seborrheic Keratosis (sk)           & 632           \\
                                          &                                        & Lichenoid Keratosis (lk)            & 6             \\ 
\cline{2-4}
                                          & Skin Appendages                        & Dermatofibroma (df)                 & 173           \\ 
\cline{2-4}
                                          & \multirow{3}{*}{Vascular}              & Hemangioma (ha)                     & 269           \\
                                          &                                        & Lymphangioma (la)                    & 10            \\
                                          &                                        & Pyogenic Granuloma (pg)             & 7             \\ 
\hline
Nonmelanocytic Indeterminate              & Keratinocytic                          & Actinic Keratosis (ak)              & 48            \\ 
\hline
\multirow{7}{*}{Nonmelanocytic Malignant} & \multirow{5}{*}{Keratinocytic}         & Basal Cell Carcinoma (bcc)           & 423           \\
                                          &                                        & Bowen's Disease (bd)                & 46            \\
                                          &                                        & Cutaneous Horn (ch)                 & 9             \\
                                          &                                        & Mammary Paget Disease (mpd)          & 13            \\
                                          &                                        & Squamous Cell Carcinoma (scc)        & 263           \\ 
\cline{2-4}
                                          & Skin Appendages                        & Dermatofibrosarcoma Protuberans (dfsp) & 4             \\ 
\cline{2-4}
                                          & Vascular                               & Kaposi Sarcoma (ks)                 & 14            \\ 
\hline
\multicolumn{3}{r|}{Total}                                                                                           & 12,345         \\
\hline
\end{tabular}
\end{table}

\section*{Data Records}
All images are in JPEG format with their corresponding label in metadata. The metadata file was generated in CSV file to process easily. The metadata contains file names and lesion classes with their detailed taxonomic identification. 

\begin{figure}[t!]
\setlength{\fboxsep}{0pt}%
\setlength{\fboxrule}{0pt}%
\centering
\includegraphics[width=\textwidth]{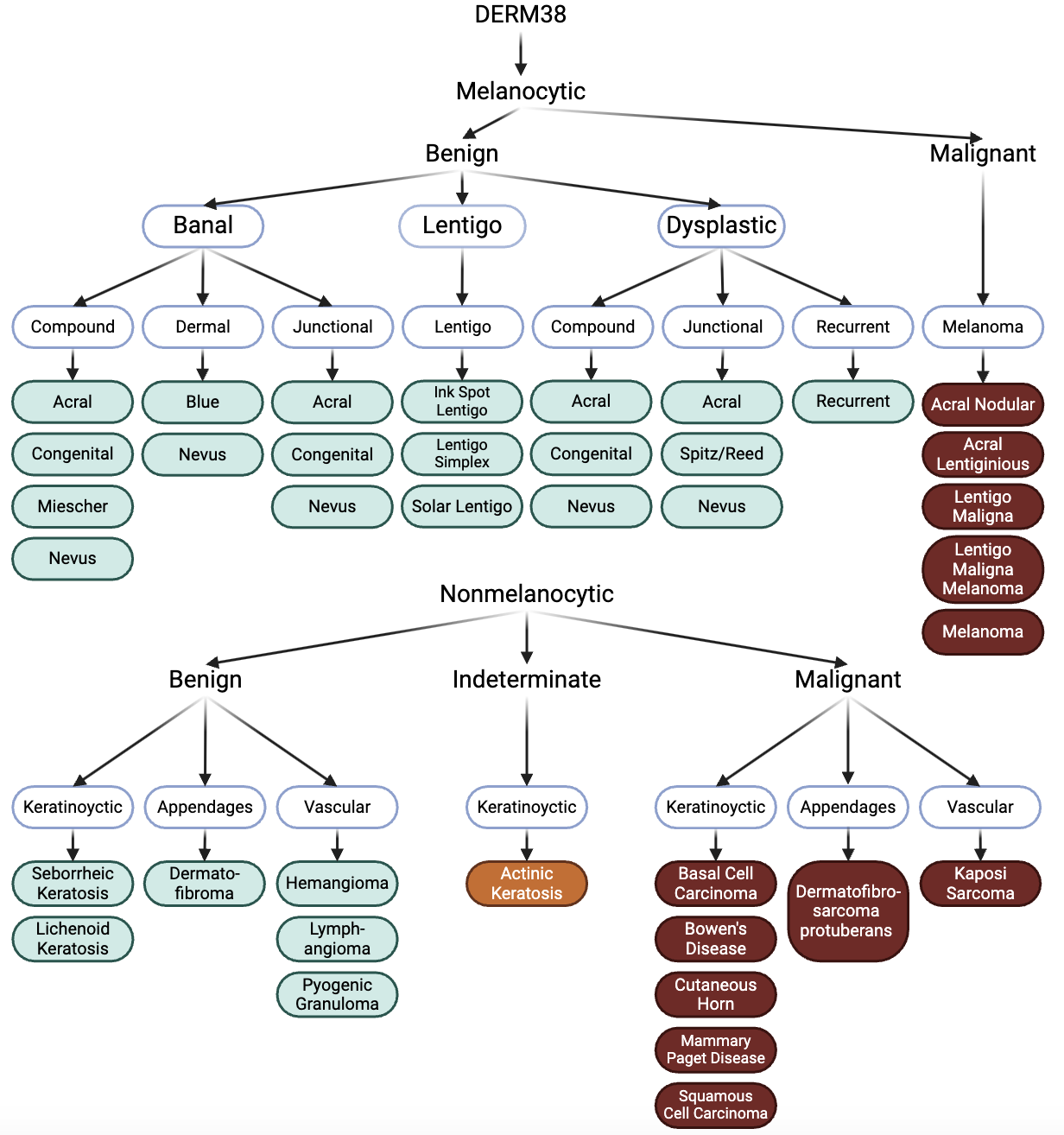}
\caption{Shows an overview of the taxonomy tree. The first level includes the melanocytic and nonmelanocytic. The second level comprises malignant and benign groups of the first level. The third level  is banal, dysplastic and lentigo for melanocytic benign, melanoma for melanocytic malignant, keratinocytic, skin appendages, vascular for nonmelanocytic benign, keratinocytic for nonmelanocytic-indeterminate, and keratinocytic, skin appendages, and vascular for nonmelanocytic malignant. The forth level is the subclasses of related to their main and super classes. Our taxonomy tree contains 5 super classes, 15 main classes, and 38 subclasses.}
\label{fig:tree}
\end{figure}

A three-level taxonomy tree following the latest dermatological classification standards \cite{scope2024international} was created to guide the use of DERM38.The benign skin lesions that are high risk of being or becoming malignant were especially classified into a separate subclass such as dysplastic nevus \cite{baigrie2018dysplastic}, congenital nevus \cite{kinsler2017melanoma}, blue nevus \cite{daltro2017atypical} and recurrent nevus that have the risk of unremoved malignant cell. The skin lesions were classified according to their specific anatomical localizations such as palm-sole (acral), face and trunk-extremities. Combined lesions with multiple pathological classification were excluded if cropping into separate images is not possible. Skin lesions were initially grouped into the two classes as melanocytic and non-melanocytic. These two classes were further divided into malignant, indeterminate and benign groups, and five super classes were created. These five super classes were then divided into 15 main classes. Lastly, sub-lesion types were then classified into 38 different subclasses. Melanocytic-benign lesions were classified into lentigo with ink spot lentigo (isl), lentigo simplex (ls), and solar lentigo (sl) subclasses, banal with dermal, compound, and junctional subclasses and dysplastic with compound, junctional, and recurrent (rn) subclasses. Dysplastic dermal nevus were added into banal dermal nevus because of their similar patterns. Common nevus subclasses (acral (an), blue (bn), Miescher (mn), congenital (cn), and spitz/reed (sn) nevus) under dermal, compound, and junctional groups were then classified into their subclasses if available. Melanoma (mel) was classified under melanocytic-malignant lesions, with five further subclasses as acral nodular melanoma (anm), acral lentiginous melanoma (alm), lentigo maligna (lm), lentigo maligna melanoma (lmm), and melanoma (mel). Nonmelanocytic-benign lesions were classified into two subclasses as keratinocytes (lichenoid keratosis (lk), seborrheic keratosis (sk)), skin appendages (dermatofibroma (df)), and vascular (hemangioma (ha), lymphangioma (la), and pyogenic granuloma (pg)). Nonmelanocytic-malignant lesions were classified into three subclasses as keratinocytes (basal cell carcinoma (bcc), Bowen’s disease (bd), cutaneous horn (ch), mammary Paget disease (mpd), and squamous cell carcinoma (scc)), skin appendages (dermatofibrosarcoma protuberans (dfsp)), and vascular (Kaposi sarcoma (ks)). In addition, actinic keratosis (ak) was classified under nonmelanocytic-indeterminate subclasses. The taxonomy tree of these classes and their sizes were shown in Fig. \ref{fig:tree} and Table \ref{tab:class_results}, respectively.

\section*{Technical Validation}

All images in this dataset were of patients applying to the dermatology department with skin complaints. All lesions were classified and reviewed according to the rules of dermatoscopic evaluation by two expert dermatologists in a consensus decision. All malignant skin lesions were all biopsy proven. The majority of these lesions consisted of follow-up images and were labeled using patient records. Nevus with over 2 years of digital dermatoscopic follow-up were annotated with no change except nevoid involution. The number of dysplastic nevus (N: 6,157) constituted a large part of the dataset. These dysplastic nevi were followed up in daily clinical practice rather than banal nevus (N:3,764). Malignant forms of benign lesions such as dfsp (N:4, counterpart of df) were also included in the dataset, even if the number of data is small. Early forms of malignant tumors such as mpd and bd which are a type of intraepidermal scc were also included. The consensus of GG and SPY experts labeled all the benign images that did not have histology or further follow-up. GG (with second Ph.D. in Basic Oncology) and SPY are two expert dermatologists with 20+ years of experience in dermatology.

\section*{Usage Notes}
The dataset provided in this study represents the initial comprehensive collection of skin lesion data from Türkiye, serving as a distinctive and important asset for both the medical and machine learning fields. The launch of this dataset represents a notable progression, especially when considering the wide range of skin types, it addresses and the comprehensive classification of skin lesions it contains. The dataset is valuable due to the inclusion of various benign subclasses that closely resemble malignant tumors, distinguishing it from current collections and presenting a novel opportunity for the advancement of advanced diagnostic algorithms.

The absence of a rich sub-classification may result in misclassification. For instance, congenital nevus (which is not categorized as a separate subclass in the published datasets) can often be mistaken as a melanoma \cite{kinsler2017melanoma}. It is also important to know that the lesion is a congenital nevus as this lesion has a high risk of evolving into a melanoma \cite{kinsler2017melanoma}. We also included uncommon lesions that are difficult to collect or annotate such as spitz/reed nevus \cite{cheng2022spectrum}. By using our detailed subclassification, this dataset enables researchers to develop and refine more reliable AI models capable of distinguishing benign from malignant ones, improving diagnostic accuracy and reducing the likelihood of misdiagnosis. The dataset includes 38 subclasses, allowing researchers to focus on more intelligent algorithms such as hierarchical learning \cite{barata2019deep}.

The dataset exhibits potential to serve as a fundamental component in the advancement of effective algorithms for identifying malignant skin lesions. The ongoing obstacle in dermatological diagnosis lies in the difficulty of differentiating between benign and malignant forms, particularly when they exhibit comparable visual characteristics. The incorporation of benign that exhibit similarities to malignant counterparts in this collection of data presents a potential for enhancing the precision of machine learning algorithms. Therefore, it is anticipated that this technology will facilitate progress in the early and accurate detection of diseases, potentially mitigating the occurrence of false positive results and ultimately enhancing patient prognoses.

The inclusion of the dataset originating from Türkiye contributes to the expansion of the worldwide data repository, which has historically been underrepresented in this research domain. The geographical specificity of this information is of great use to academics who seek to create diagnostic models that are both resilient and efficient across various ethnicities and geographic areas. Additionally, it functions as a valuable resource for doctors who aim to comprehend the diversity in lesion, which can be impacted by various factors such as regional environmental conditions and genetic predispositions.

Clinicians are encouraged to employ this dataset as a visual resource for the identification of skin lesions, while researchers are encouraged to utilize it for comparative analyses with other datasets specific to the location. This dataset is of particular utility to machine learning practitioners for the purposes of training and verifying classification algorithms. The comprehensive depiction of complex instances within the dataset offers a rigorous platform for evaluating algorithms intended to distinguish between benign and malignant tumors.

In brief, this dataset serves to enhance the existing body of data pertaining to skin lesions, while also paving the way for the development of sophisticated diagnostic instruments. We anticipate that this tool will be a helpful resource in enhancing the precision and effectiveness of skin cancer detection and diagnosis among various demographic groups.

In this dataset, there are 38 subgroup lesions. Brief descriptions of these subgroups presented below can be useful for AI researchers:\\

I. Melanocytic lesions: Originates from melanocytes, often pigmented, shades of brown, black, tan, blue. Reflecting varying concentrations and depth of melanin within the skin.
   \begin{enumerate}
         \item Benign: Non-cancerous melanocyte-origin moles, generally harmless.
         \begin{enumerate}
         \item Dysplastic Nevus: Also known as atypical or Clark’s nevus. They are characterized by histologic features, and they may appear small and banal clinically. Dysplastic nevi demonstrate the following clinical features: usually > 5 mm, irregular borders, some with a pigmented and erythematous rim, and variegated pigmentation with a mixture of pink, light and dark brown colors \cite{carrera2016discriminating,suh2018dysplastic}. This subclass contains compound nevus with acral and congenital subclasses, junctional nevus with acral and spitz/reed subclasses, and recurrent nevus.
         \item Banal Nevus: A banal nevus, commonly known as a common mole, is a typical, non-cancerous skin mole that usually appears as a small, regular, round, or oval spot. Brown, tan, or pink, and may be either flat or slightly raised. This subclass contains dermal nevus with blue subclass, compound nevus with acral, Miescher, and congenital subclasses, and junctional nevus with acral and spitz/reed subclasses.
         \item Lentigo: Pigmented and uniform. It is related to increasing of melanin expression by melanocytes. This subclass contains ink spot lentigo, lentigo simplex, and solar lentigo.
         \begin{enumerate}
            \item Compound (cn): Common mole, partially flat and partially elevated. Mix of dermal and epidermal melanocytes. 
            \item Dermal (dn): Common mole. Raised bump, flesh-colored or brown mole. Located within the dermis. Generally benign. Unna nevus is included.
            \item Junctional (jn): Flat, typically dark, or brown mole. Occurs at epidermis-dermis junction. 
            \item Recurrent nevus (rn): Incompletely removed nevus. Mole regrowth at previously excised site. Typically, benign. Can be confused with melanoma.
            \begin{enumerate}
         \item Acral nevus (an): Mole on palms or soles. Usually benign and characterized by their distinctive location on the body where the skin is thicker.
         \item Blue nevus (bn): Deeply pigmented mole, blue-black color. Originates from deep dermal melanocytes. Usually, benign. 
         \item Congenital nevus (con): Mole present at birth and can grow proportionally with the child. Varies in size; potentially large. Increased melanoma risk later in life. Nevus spilus is included.
         \item Ink Spot Lentigo (isl): Small, dark brown to black spots, resembling ink spots.
         \item Lentigo Simplex (ls): Small, a precursor to junctional nevus. Small, flat, and typically darker spot. Also known as simple lentigo.  
         \item Miescher nevus (mn): Reveal pseudo network around hair follicles. Commonly benign, typically brown.
         \item Solar Lentigo (sl): Larger, a precursor to seborrheic keratosis. Sun exposed.
         \item Spitz/Reed nevus (sn): Raised, pink, red or brown mole. Often mistaken for melanoma. Typically, benign in children. Sometimes, shows a starburst pattern. 
       \end{enumerate}
        \end{enumerate}
       \end{enumerate}
         \item Malign: Cancerous growth from melanocytes. High risk of spreading.
         \begin{enumerate}
         \item Melanoma: Aggressive skin cancer. Arises from melanocytes. High risk of metastasis. 
         \begin{enumerate}
         \item Acral nodular melanoma (anm): Nodular melanoma form on poles and soles.
         \item Acral lentiginous melanoma (alm): A rare subtype of melanoma. Lentiginous form on poles and soles.
         \item Lentigo maligna (lm): Larger, a precursor to lentigo maligna melanoma. Slow-growing skin cancer. Appears as a flat, blotchy patch. Early form of melanoma. Located on sun-exposed areas of the skin.
         \item Lentigo maligna melanoma (lmm): Various number of colors, especially blue or black that means the melanoma cells have reached to the deeper layers of skin.
         \item Melanoma (mel): Melanoma can appear as a new dark spot on the skin or from an existing mole that changes in color, size, or feel. It can spread quickly to other parts of the body and is critical to treat early.
       \end{enumerate}
       \end{enumerate}
       \end{enumerate}

II. Nonmelanocytic lesions: Uncontrollable growth pattern from keratinocytes, fibroblasts, or vascular cells.
\begin{enumerate}
         \item Benign: Non-cancerous growths from keratinocytes, fibroblasts, or vascular cells.
         \begin{enumerate}
         \item Keratinocytic:
          \begin{enumerate}
\item Seborrheic keratosis (sk): Appears as a waxy, wart-like, often brown or black growth. Commonly has a rough texture. Stucco keratosis is included which is located on legs. 
\item Lichenoid keratosis (lk): Inflammatory and regressing on solar lentigo and seborrheic keratosis.
\end{enumerate}
\item Skin appendages:
\begin{enumerate}
\item Dermatofibroma (df): Firm, benign skin nodule. Common on legs or arms. Typically, brownish, and harmless.
\end{enumerate}
\item Vascular:
\begin{enumerate}
\item Hemangioma (ha): Bright red, raised birthmark. Made of clustered blood vessels. Common in infants, usually fades.
\item Lymphangioma (la): Rare, soft, often translucent mass. Appears as a swelling on the skin or mucous membranes. Typically seen in infants.
\item Pyogenic granuloma (pg): Small, reddish, raised lesion. Prone to bleeding and rapid growth. Often appears after skin injury.
\end{enumerate}
\end{enumerate}
\item Indeterminate: Not exactly categorized as benign or malignant. This does not imply that these lesions are malignant.
\begin{enumerate}
\item Keratinocytic:
\begin{enumerate}
\item Actinic keratosis (ak): Rough, scaly skin patch. Sun-induced, precancerous. Common on sun-exposed areas. 20\% malignancy risk.
\end{enumerate}
\end{enumerate}
\item Malign: Cancerous growths, and can originate from keratinocytes, vascular, or connective tissue cells.
\begin{enumerate}
\item Keratinocytic:
\begin{enumerate}
\item Basal cell carcinoma (bcc): Appears as a shiny, pearly bump or a flat, scar-like lesion. Most common on sun-exposed areas. Slow growing rarely metastasizes. 
\item Bowen’s disease (bd): Early form of scc. Appears as a persistent, red, scaly patch. Can resemble eczema or psoriasis. Often occurs on sun-exposed skin. 
\item Cutaneous horn (ch): Hard, protruding growth resembling an animal horn. Composed of compacted keratin mostly located on the face, ears, and other sun exposed areas. It may be associated with benign, premalignant, and malignant lesions. Often benign, base may have underlying malignancy. Only early form of scc is included.
\item Mammary Paget’s Disease (mpd): Early form of scc and a type of breast cancer. mpd need to be diagnosis before it converts to nodular form.
\item Squamous cell carcinoma (scc): Common, appears as a firm, red nodule, or a flat lesion with a scaly, crusted surface. Commonly develops in sun-exposed areas. Can be ulcerative and may bleed. 
\end{enumerate}
\item Skin appendages:
\begin{enumerate} 
\item Dermatofibrosarcoma protuberans (dfsp): Rare, appears as a deep-seated, firm lump under the skin. Often starts as a small, painless nodule that gradually enlarges. Typically develops on the trunk or limbs.
\end{enumerate}
\item Vascular:
\begin{enumerate} 
\item Kaposi sarcoma (ks): Presents as purple, red, or brown blotches or tumors on the skin. It is originated from vessel’s endothelial cells.
\end{enumerate}
\end{enumerate}
\end{enumerate}



\bibliographystyle{unsrt}
\bibliography{bibliography.bib}

\section*{Acknowledgements}
Abdurrahim Yilmaz has been funded by the President’s PhD Scholarships of Imperial College London. 

All figures are created with BioRender.com.

\section*{Author contributions statement}

Conceptualization, A.Y., S.P.Y., and G.G.; methodology, A.Y., G.G., and B.T.; software, A.Y.; validation, G.G., S.P.Y., and B.T.; formal analysis, A.Y.; investigation, G.G. and S.P.Y.; resources, G.G. and S.P.Y.; data curation, A.Y., G.G., S.P.Y., and B.T.; writing---original draft preparation, A.Y., and B.T.; writing---review and editing, A.Y. and G.G.; visualization, A.Y. All authors have read and agreed to the published version of the manuscript.

\section*{Competing interests}
There is no competing interest.

\end{document}